\documentclass[aps,prl,reprint,groupedaddress]{revtex4-1}

\usepackage{graphicx}
\usepackage{bm}
\usepackage{amssymb}
\usepackage{amsmath}
\usepackage{hyperref}
\usepackage{comment}
\usepackage{mathtools}
\usepackage{color}
\hypersetup{colorlinks=true, citecolor=blue,linkcolor=blue,urlcolor=blue}

\begin{document}


\title{
Rigorous Results for the Ground States of the Spin-2 Bose-Hubbard Model
}



\author{Hong Yang}
\author{Hosho Katsura}
\affiliation{
Department of Physics, Graduate School of Science, The University of Tokyo,
Hongo, Tokyo 113-0033
}


\date{February 7, 2019}

\begin{abstract}
We present rigorous and universal results for the ground states of the $f=2$ spinor Bose-Hubbard model. 
The model includes three two-body on site interaction terms, 
two of which are spin dependent while the other one is spin independent.
We prove that, depending only on the coefficients of the two spin dependent terms,
the ground state exhibits maximum or minimum total spin or SU(5) ferromagnetism.
Exact ground-state degeneracies and the forms of ground-state wave function are also determined in each case.
All these results are valid regardless of dimension, lattice structure, or particle number.
Our approach takes advantage of the symmetry of the Hamiltonian and employs mathematical tools 
including the Perron-Frobenius theorem and the Lie algebra $\mathfrak{so}(5)$. 

\end{abstract}


\maketitle

{\it Introduction.}---Ultracold atoms in optical lattices provide a unique playground for studying quantum many-body systems experimentally~\cite{Morsch_2006, Bloch_2008, Lewenstein2012, Krutitsky_2016}.
In particular, systems of bosonic alkali atoms with hyperfine spin $f$ 
have received considerable attention, as they can give rise to a variety of exotic phases~\cite{ueda2010fundamentals, Kawaguchi_Ueda_2012, Ueda_RMP_2013}. 
Such systems are well described by the spinor Bose-Hubbard model~\cite{Demler_Zhou_2002,Imambekov_2003,Lewenstein2012}, which is a discrete version of the model for condensates~\footnote{When all atoms are trapped in one single potential well, they form a continuous condensate~\cite{Koashi2000, Ciobanu2000, Lewenstein2012}. }.
Most previous theoretical studies on lattice systems were based on a mean-field treatment of the original model~\cite{Zhou_Snoek_2003, Hou_Ge_2003, Snoek_Zhou_2004, 
Imambekov_2004, Tsuchiya_2004, Kimura_2005, Krutitsky_2005, Zhou_Semenoff_2006, Snoek_Song_2009} or a mapping to the effective spin model (perturbation on hopping)~\cite{Imambekov_2003, Zhou_Snoek_2003, Yip_2003, DeChiara_2011, eckert2007ultracold, barnett2006classifying}. 
This is in contrast to the continuous case, where many results beyond mean-field have been obtained theoretically~\cite{Koashi2000,Ueda2002,uchino2008dynamical,Ho1999,
eisenberg2002polarization,Tasaki_polar_2013}. 
On the other hand, very few solid results are available for discrete lattice systems~\cite{Katsura2013, Kunimi_Saito_2015}.
In particular, almost nothing is known rigorously about the $f=2$ case.

\smallskip


In this Letter, we prove universal theorems
about the ground-state properties of the \mbox{spin-2} Bose-Hubbard Hamiltonian.  The model in Eq.\ (\ref{Hamiltonian}) is characterized by the spin-dependent interaction constants $c_1$ and $c_2$. We precisely determine the total spin and the degeneracy of the ground states for the following three cases: (i) $c_1<0$ and $c_2 \geqslant 5c_1$, (ii) $c_1=0$ and $c_2<0$, and (iii) $c_1=c_2=0$. 
In case (i), the system has SO(3) symmetry and the ground state exhibits saturated ferromagnetism. 
In case (ii), the symmetry is promoted to SO(5) 
and the ground state has total spin $F_{\text{tot}}=0$ or $2$, depending on the parity of the total particle number. The symmetry is further enhanced to SU(5) in case (iii), leading to a ground-state degeneracy that grows as $N^4$ for large particle number $N$.
It is worth comparing
with the previous results for the spin-1 Bose-Hubbard model~\cite{Katsura2013}. 
The spin-2 Hamiltonian has one more spin-dependent interaction term with SO(5) symmetry
that does not have a counterpart in the spin-1 case, 
which makes the phase diagram richer.


\smallskip

Our results are consistent with the phase diagram of spin-2 continuous condensates predicted by mean-field theory~\cite{Koashi2000,Ciobanu2000,Lewenstein2012}.
Moreover, our Theorem 2 suggests that the model with $c_1=0$ and $c_2<0$ can be studied by quantum Monte Carlo simulations without a sign problem~\cite{
[{A sign problem may occur in boson systems with spin-dependent interaction.
The one-dimensional spin-1 Bose-Hubbard model with antiferromagnetic interactions was studied using quantum Monte Carlo simulations in }]
Apaja_2006}.


\smallskip

{\it Hamiltonian.}---We consider a system of $N$ spinor bosons with $f=2$ on a finite set of sites $\Lambda$, where $N$ is arbitrary and fixed.
We use Latin letters $i$ and $j$ to denote sites
and Greek letters $\alpha$ and $\beta$ 
to represent spin states,
i.e., $i,j \in \Lambda$ and $\alpha,\beta \in \{+2,+1,0,-1,-2\}$.
The creation and annihilation operators at
site $i$ with spin $\alpha$ are written as
${\hat a}^\dagger_{i, \alpha}$ and ${\hat a}_{i, \alpha}$, respectively.
We denote the number operators by $\hat{n}_{i,\alpha} := {\hat a}^\dagger_{i, \alpha} {\hat a}_{i, \alpha}$, $\hat{n}_i := \sum_\alpha \hat{n}_{i,\alpha}$, 
and $\hat{N}_{\alpha} := \sum_i \hat{n}_{i,\alpha}$, and the $z$ component of the spin operator by
$\hat{F}^z_i := \sum_{\alpha, \beta} {\hat a}^\dagger_{i, \alpha} F^z_{\alpha, \beta} {\hat a}_{i,\beta}$,
where $F^z_{\alpha, \beta}$ is the spin matrix for spin-2. 
Similarly, we have $\hat{F}^x_i$ and $\hat{F}^y_i$.
We define $\hat{\bm{F}}_i :=( \hat{F}^x_i, \hat{F}^y_i, \hat{F}^z_i )$ and $\hat{\bm{F}}_{\text{tot}} := \sum_i \hat{\bm{F}}_i$, and write the eigenvalues of 
$(\hat{\bm{F}}_{\text{tot}})^2$ and
$\hat{F}^z_{\text{tot}}$ as $F_{\text{tot}} (F_{\text{tot}} +1)$ and
$F^z_{\text{tot}}$, respectively.  
We also define the singlet creation and annihilation operators as $\hat{S}_{i, +} := \sum_{\alpha=-2}^2 (-1)^{\alpha}{\hat a}^{\dagger}_{i,\alpha} {\hat a}^{\dagger}_{i,-\alpha}/2$
and $\hat{S}_{i, -} := \hat{S}_{i, +}^\dagger$.
Acting with $\hat{S}_{i,+}$ ($\hat{S}_{i,-}$) creates (annihilates) a two-body singlet at site $i$. 

\smallskip

The following set of orthonormal states
\begin{equation}
|\Phi_{\bm{m}}\rangle:= \frac{1}{\sqrt{\prod_{i,\alpha}(n_{i,\alpha} !)}} \left\{ \prod\limits_{i, \alpha} (\hat{a}_{i, \alpha}^{\dagger})^{n_{i, \alpha}} \right\} |\text{vac}\rangle \label{basis}
\end{equation}
serves as a basis 
of the Hilbert space $\mathcal{H}$. 
Here, $|\text{vac}\rangle$ stands for the vacuum and $\bm{m}=(n_{i,\alpha}) \in \mathcal{I}$ is 
a set of non-negative integers, where $\mathcal{I}$ is a set of ${\bm m}$ that satisfies $\sum_{i,\alpha}n_{i,\alpha}=N$.

\smallskip

The Hamiltonian of the spin-2 Bose-Hubbard model~\cite{Koashi2000,Lewenstein2012}
is
\begin{eqnarray}
  \hat{H}=&&-\sum_{ i\neq j, \alpha} t_{i, j} \hat{a}^{\dagger}_{i,\alpha} \hat{a}_{j,\alpha}
            + \sum_{i}V_i \hat{n}_i + \frac{c_0}{2}\sum_{i}\hat{n}_i(\hat{n}_i-1) \nonumber\\
          &&+ \frac{c_1}{2} \sum_{i}\Big[(\hat{\bm{F}_i})^2 - 6\hat{n}_i\Big]
            + \frac{2c_2}{5} \sum_{i}\hat{S}_{i, +}\hat{S}_{i, -}~.  \label{Hamiltonian}
\end{eqnarray}
Here $V_i \in \mathbb{R}$ is the single-particle potential at site $i$. The constants $c_0$, $c_1$ and $c_2$ are real coefficients for the two-body interactions, where the $c_1$ and $c_2$ terms are spin dependent. The $c_2$ term favors (disfavors) singlet pairs when $c_2 > 0$ ($c_2 < 0$). We assume that $t_{i,j}=t_{j,i} \geqslant 0$ for all $i,j \in \Lambda$ and the whole lattice $\Lambda$ is connected via nonzero $t_{i,j}$. 

\smallskip

In addition to the global ground states in the whole Hilbert space $\mathcal{H}$,
symmetry of the Hamiltonian enables us
to discuss the local ground states in Hilbert subspaces.
The Hamiltonian ${\hat H}$ is invariant under rotation in spin space, which implies that ${\hat H}$ has at least SO(3) symmetry, yielding $[\hat{H}, \hat{F}^{x,y,z}_{\text{tot}}] = 0$.
Since $\hat{F}_{\text{tot}}^z$ is conserved,
$\mathcal{H}$ splits into subspaces labeled by $F^z_{\text{tot}}=M$. 
We shall show later that the symmetry is promoted to SO(5) when $c_1=0$ and $c_2 \neq 0$. In this case, $\mathcal{H}$ splits into smaller subspaces labeled by two indices $P:=N_1-N_{-1}$ and $Q:=N_2-N_{-2}$.

\smallskip

Now we introduce some more notations. Define $\mathcal{H}_A$ as a subspace of $\mathcal{H}$ by
$\mathcal{H}_A := \{ |\psi\rangle \in \mathcal{H} \mid \hat{A} |\psi\rangle = A |\psi\rangle \}$. 
Similarly, we have $\mathcal{H}_B$ for ${\hat B}$. 
The intersection of $\mathcal{H}_{A}$ and $\mathcal{H}_{B}$ is denoted as $\mathcal{H}_{A,B}$.  
Define $\mathcal{I}_A$ as a subset of $\mathcal{I}$ by
$\mathcal{I}_A := \{ \bm{m} \in \mathcal{I} \mid \hat{A} |\Phi_{\bm{m}}\rangle = A |\Phi_{\bm{m}}\rangle  \}$. 
Operators $\hat{A}$ and $\hat{B}$ can be $\hat{M}$ $(:= \hat{F}^z_{\text{tot}})$, $\hat{P}$ $(:= \hat{N}_{1}-\hat{N}_{-1})$, $\hat{Q}$ $(:=\hat{N}_{2}-\hat{N}_{-2})$, or $\hat{N}_\alpha$ in the following.

\smallskip
 
Now we state our main theorems. 

\smallskip

\textit{Theorem 1.---}If $c_1<0$ and $c_2 \geqslant 5c_1$, the local ground state $|\Psi^{\text{GS}}_M\rangle$ in $\mathcal{H}_M$ is unique and can be written as
\begin{equation}
|\Psi^{\text{GS}}_M\rangle = \sum\limits_{\bm{m} \in \mathcal{I}_M} C_{\bm{m}} |\Phi_{\bm{m}}\rangle, \label{Thm1}
\end{equation}
with $C_{\bm{m}} > 0$, and has the maximum possible total spin $F_{\text{tot}}=2N$ (saturated ferromagnetism). Each local ground state $|\Psi^{\text{GS}}_M\rangle$ has energy 
independent of $M$ and hence is the global ground state in $\mathcal{H}$ as well. Thus the ground-state degeneracy is $4N+1$.

\smallskip

The following proposition is a special case of Theorem~1 where $|\Psi^{\text{GS}}_M\rangle$ can be written more explicitly. 

\smallskip

\textit{Proposition 1.---}If $c_1=-c_0/4 < 0$ and $c_2\geqslant0$,
the ground state $|\Psi^{\text{GS}}_{M=2N}\rangle$ in $\mathcal{H}_{M=2N}$ is unique
and can be written as
\begin{eqnarray}
&&|\Psi^{\text{GS}}_{M=2N}\rangle  = (\hat{b}^{\dagger}_{2})^{N} |\text{vac}\rangle,
\label{Prop1}
\end{eqnarray}
where $\hat{b}^{\dagger}_2 = \sum_i \varphi_0(i) \hat{a}^{\dagger}_{i,2}$.
Here, $\varphi_0(i) >0$ ($i \in \Lambda$) is the 
spatial wave function of single-particle ground state of the hopping term and the on-site potential term.
Clearly $|\Psi^{\text{GS}}_{M=2N}\rangle$ has the maximum total spin $F_{\text{tot}}=2N$.
Ground states in other subspaces can be obtained as 
$|\Psi^{\text{GS}}_{M'}\rangle \propto (\hat{F}_{\text{tot}}^-)^{2N-M'} |\Psi^{\text{GS}}_{M=2N}\rangle$.

\smallskip

\textit{Theorem 2.---}If $c_1=0$ and $c_2<0$, the local ground state $|\Psi^{\text{GS}}_{P,Q}\rangle$ in $\mathcal{H}_{P,Q}$ is unique and can be written as
\begin{equation}
|\Psi^{\text{GS}}_{P,Q}\rangle =\sum\limits_{\bm{m}\in\mathcal{I}_{P,Q}} D_{\bm{m}} (-1)^{ (\hat{N}_{+1} + \hat{N}_{-1})/2} |\Phi_{\bm{m}}\rangle, \label{Thm2}
\end{equation}
with $D_{\bm{m}} > 0$. The local ground-state energy 
in each $\mathcal{H}_{P,Q}$ is a function only of $\Gamma:= |P|+|Q|$.
We denote this energy by $E^{\text{GS}}_\Gamma$.
Their energy-level ordering is 
$E^{\text{GS}}_\Gamma < E^{\text{GS}}_{\Gamma+1}$ if $N-\Gamma$ is even, 
while $E^{\text{GS}}_\Gamma = E^{\text{GS}}_{\Gamma+1}$ if $N-\Gamma$ is odd. 
Thus the global ground state has total spin $F_{\text{tot}}=0$ and is unique 
if total particle number $N$ is even, 
while it has $F_{\text{tot}}=2$ and is fivefold degenerate if $N$ is odd. 

\smallskip

Note that $\mathcal{H}_{P,Q} \subset \mathcal{H}_{M=P+2Q}$. When $c_1=0$ and $c_2<0$, 
due to the SO(5) symmetry of the $c_2$ term (to be shown in the Proof of Theorem 2),
the Hamiltonian conserves two quantities $P = N_1 - N_{-1}$ and $Q = N_2 - N_{-2}$.
Nevertheless, the energy-level ordering is determined by only one quantum number $\Gamma$.
The fact that the ground state tends to be a singlet is consistent with what one would expect from the $c_2 (>0)$ term which favors spin-singlet pairs. 

\smallskip

\textit{Theorem 3.---}If $c_1 = c_2 = 0$, the local ground state $|\Psi^{\text{GS}}_{N_2,\ldots,N_{-2}}\rangle$ in $\mathcal{H}_{N_2,\ldots,N_{-2}} (:= \bigcap^2_{\alpha=-2} \mathcal{H}_{N_\alpha})$ is unique and can be written as
\begin{equation}
|\Psi^{\text{GS}}_{N_2,\ldots,N_{-2}}\rangle =\sum\limits_{\bm{m}\in \mathcal{I}_{N_2,\ldots,N_{-2}}} G_{\bm{m}} |\Phi_{\bm{m}}\rangle, \label{Thm3}
\end{equation}
with $G_{\bm{m}} > 0$.
The local ground-state energy is independent of $N_2,\ldots, N_{-2}$.
Thus each $|\Psi^{\text{GS}}_{N_2,\ldots,N_{-2}}\rangle$ is also the global ground state in $\mathcal{H}$,
and the 
ground-state degeneracy is $\binom{N+4}{4}=(N+4)!/(N!4!)$.

\smallskip

Note that $\mathcal{H}_{N_2,\ldots,N_{-2}} \subset \mathcal{H}_{M=2(N_2-N_{-2})+N_1-N_{-1}}$. Because of the absence of the spin-dependent interaction, the Hamiltonian in this case has SU(5) symmetry and conserves the particle number of each spin state. We can say that the ground states exhibit ``SU(5) ferromagnetism".

\smallskip

\begin{figure}[b]
\includegraphics[width=0.37\textwidth]{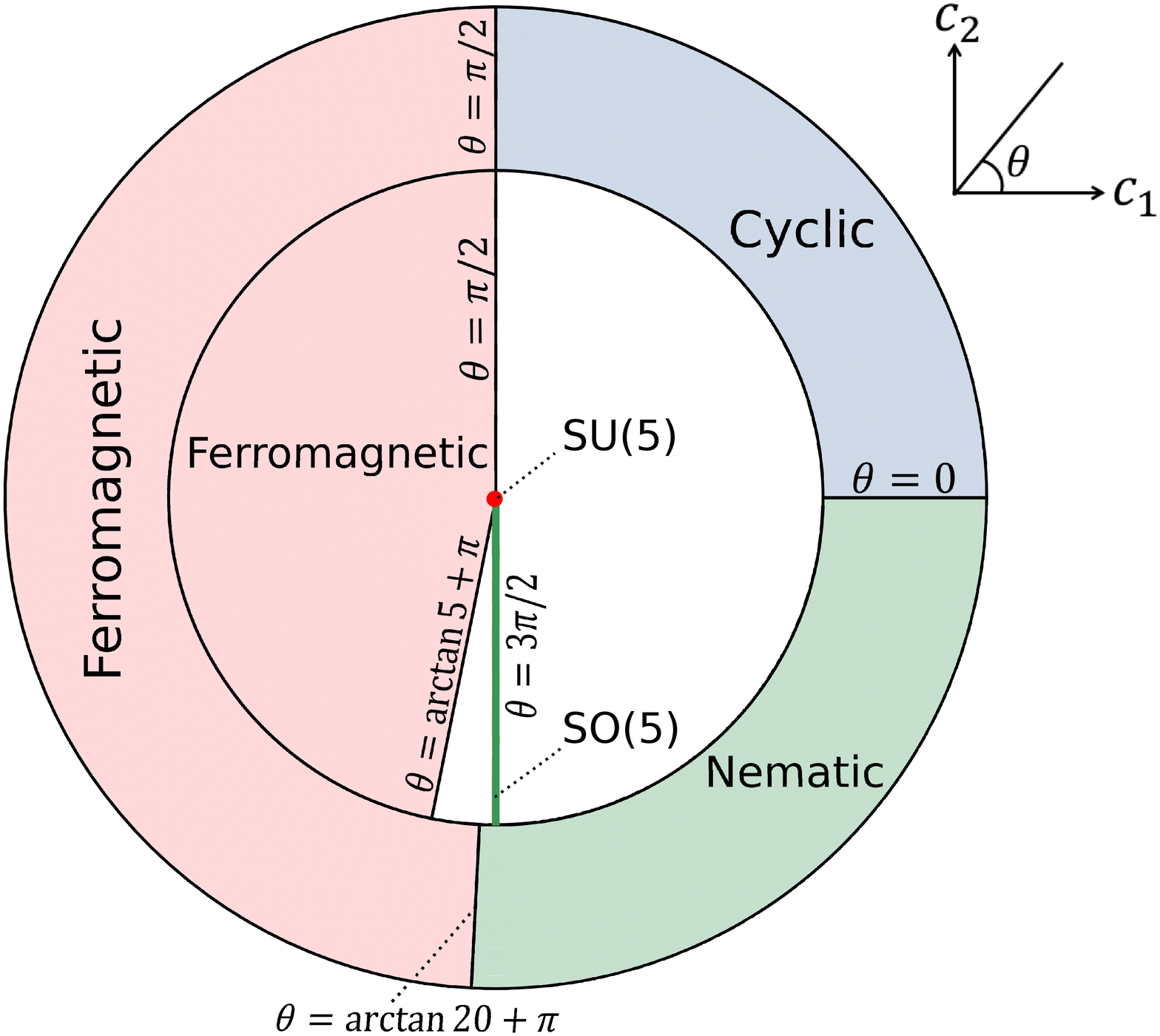}
\caption{\label{phase_diagram} Regions of our results for discrete lattices (inner circle) and the mean-field ground-state phase diagram of spin-2 Bose-Einstein continuous condensates (outer circle)~\cite{ueda2012bose,Koashi2000,Ciobanu2000,Lewenstein2012,barnett2006classifying}. 
In the small hopping limit of a homogenous lattice system, the outer circle is also the phase diagram of a Mott insulator state with one boson per site~\cite{Lewenstein2012,barnett2006classifying}.
Here, $\tan \theta= c_2/c_1$. 
At the mean-field level, ferromagnetic, nematic, and cyclic phases are characterized by $(\langle \hat{\bm{F}}\rangle\neq 0$, $\langle \hat{S}_{+}\hat{S}_{-} \rangle=0)$, $(\langle \hat{\bm{F}}\rangle= 0$, $\langle \hat{S}_{+}\hat{S}_{-} \rangle \neq 0)$, and $(\langle \hat{\bm{F}}\rangle= 0$, $\langle \hat{S}_{+}\hat{S}_{-} \rangle=0)$, respectively.
}
\end{figure}

The above three theorems concern the ground-state magnetic properties and degeneracies. 
In Fig.~\ref{phase_diagram}, the regions of these three theorems are shown together 
with the mean-field phase diagram of spin-2 condensates.


\smallskip

{\it Proofs.}---It is worth noting that, if $\hat{H}$ is expanded in terms of bosonic operators $\hat{a}^{\dagger}$ and $\hat{a}$, 
the coefficient of each term implies the matrix element $\langle \Phi_{\bm{m}} | \hat{H} | \Phi_{\bm{m}'} \rangle$.
As a simple example, the hopping term always
results in nonpositive off-diagonal matrix elements because $-t_{i,j} \leqslant 0$. 

\smallskip

\textit{Proof of Theorem 1.---}We first consider a single-site model in which $N$ particles sit on the same site $q \in \Lambda$~\footnote{The single-site model is essentially the same as continuous condensates with single-mode approximation applied~\cite{ueda2010fundamentals, Koashi2000, uchino2008dynamical}.}. 
The Hamiltonian of the model can be obtained by taking $t_{i,j}=0$ for all $i,j$ in Eq.~(\ref{Hamiltonian}). 
Let us first prove the following lemma. 

\smallskip

\textit{Lemma.---}Every local ground state $| \widetilde{\Psi}^{\text{GS}}_M \rangle$ of the single-site model has a total spin $F_{\text{tot}}=2N$.

\smallskip

\textit{Proof of Lemma.---}Recall the SO(3) symmetry of the Hamiltonian. 
Without hopping, we then see that all terms in the Hamiltonian commute with each other, which allows us to explicitly write down the energy eigenvalues of the system (see \cite{Koashi2000,Ueda2002,uchino2008dynamical} or 
Supplemental Material~\footnote{\label{footnote1}See Supplemental Material at \href{http://link.aps.org/supplemental/10.1103/PhysRevLett.122.053401}{http://link.aps.org/supplemental/10.1103/PhysRevLett.122.053401}
for how to understand the SO(5) symmetry of $c_2$ term and how to find its eigenstates 
and energy eigenvalues from the highest-weight representation of $\mathfrak{so}(5)$, which includes~\cite{Ueda2002,georgi1999lie,zee2016group}.}),
\begin{eqnarray}
E = &&V_q N + \frac{c_0}{2}N(N-1)
    + \frac{c_1}{2}[ F_{\text{tot}}(F_{\text{tot}}+1)-6 N ] \nonumber\\
 &&+ \frac{c_2}{10} (N^2+3N-v^2-3v),
\label{E(one_site)}
\end{eqnarray}
where $v$ is the number of bosons that do not form (two-particle) singlets. 
To minimize $E$ in $\mathcal{H}_M$, note that $c_1 <0$ and $0 \leqslant F_{\text{tot}}=F_q \leqslant 2v$.
A simple analysis yields $F_{\text{tot}}=2v=2N$ 
for every local ground state $| \widetilde{\Psi}^{\text{GS}}_M \rangle$.

\smallskip

Theorem~1 can now be proved in two separate regions.

\smallskip

(1) $\{c_1<0, 5c_1 \leqslant c_2 \leqslant 0\} $: By directly expanding $\hat{H}$ in terms of $\hat{a}^{\dagger}$ and $\hat{a}$'s,
one can easily find that $\forall \bm{m} \neq \bm{m}'$, $\langle \Phi_{\bm{m}} | \hat{H} | \Phi_{\bm{m}'} \rangle \leqslant 0$ is always true.
Because of the SO(3) symmetry, in the 
basis $\{| \Phi_{\bm{m}}\rangle\}$,
the matrix of $\hat{H}$ is real symmetric and block diagonal with respect to $M$. 
Within each $\mathcal{H}_M$, all possible configurations (distributions of particles on $\Lambda$, regardless of their spins) are connected via hopping,
and all possible spin states are connected via spin-dependent interactions $c_1$ and $c_2$ terms. 
Therefore, for each block of $\hat{H}$, we can apply the Perron-Frobenius theorem \cite{
[{Let $M$ be a finite-dimensional real symmetric matrix with the following properties:
(a) $M_{ij}\leqslant0$ for any $i\neq j$;
(b) all $i\neq j$ are connected via nonzero matrix elements [i.e., for any $i\neq j$, $\exists (i_1, \cdots,i_n)$ with $i_1=i$, $i_n=j$, such that $M_{i_1,i_2}M_{i_2,i_3} \cdots M_{i_{n-1},i_n}\neq0$].
Then the lowest eigenvalue of $M$ is nondegenerate and all the components of the corresponding eigenvector can be taken to be strictly positive.
This is a corollary of the Perron-Frobenius theorem. For proof, see }]
Tasaki1998},
which implies that the local ground state $|\Psi^{\text{GS}}_M\rangle$ in $\mathcal{H}_M$ is unique and can be written as Eq.~(\ref{Thm1}). 
Since $(\hat{\bm{F}}_{\text{tot}})^2$ commutes with both ${\hat H}$ and ${\hat M}$, each $|\Psi^{\text{GS}}_M\rangle$ must be an eigenstate of $(\hat{\bm{F}}_{\text{tot}})^2$. 
To determine the total spin of $|\Psi^{\text{GS}}_M\rangle$, consider the overlap between $|\Psi^{\text{GS}}_M\rangle$ and $| {\widetilde \Psi}^{\text{GS}}_M \rangle$. Since the Perron-Frobenius theorem also applies to the single-site model and implies that the ground state $| {\widetilde \Psi}^{\text{GS}}_M \rangle$ has an expansion similar to Eq. (\ref{Thm1}) with $C_{\bm m} \geqslant 0$, we have 
$\langle \widetilde{\Psi}^{\text{GS}}_M | \Psi^{\text{GS}}_M \rangle \neq 0$. 
This means that the total spin of $|\Psi^{\text{GS}}_M\rangle$ is the same as that of $| {\widetilde \Psi}^{\text{GS}}_M \rangle$. It then follows from the Lemma that $|\Psi^{\text{GS}}_M\rangle$ has the total spin $F_{\text{tot}}=2N$. 

\smallskip

(2) $\{ c_1<0, c_2>0 \}$:
In this region, we cannot apply the Perron-Frobenius theorem in the basis Eq.~(\ref{basis}), 
because the off-diagonal matrix elements of ${\hat H}$ take both positive and negative values. 
Instead, we use the min-max theorem~\footnote{{Let $\hat{A}$ and $\hat{B}$ be two Hermitian operators on a finite-dimensional Hilbert space $\mathcal{H}$ satisfying
$\langle \alpha |(\hat{A}-\hat{B})|\alpha\rangle\geqslant0$, $\forall |\alpha\rangle\in\mathcal{H}$. (i.e., $\hat{A}-\hat{B}$ is positive semidefinite.)
Let $a_i$ and $b_i$ be the $i$ th eigenvalues of $\hat{A}$ and $\hat{B}$, respectively. $a_i$ and $b_i$ are arranged so that $a_1\leqslant a_2\leqslant\cdots$, $b_1\leqslant b_2\leqslant\cdots$. Then the min-max theorem implies that $a_i \geqslant b_i$, $\forall i$.}}.
Define $\hat{H}_a := \hat{H}(c_1<0, 5c_1 \leqslant c_2 \leqslant 0)$ and $\hat{H}_b := \hat{H}(c_1<0, c_2>0)$.
Also in each $\mathcal{H}_M$, define $E_{a,0}$ and $E_{a,1}$ as the energies of the local ground state and the first local excited state of $\hat{H}_a$, respectively.
Similarly, we have $E_{b,0}$ and $E_{b,1}$.
(If there is degeneracy in the ground state, then $E_{b,1} = E_{b,0}$.) 
The local ground state of $\hat{H}_a$, as proved above, is a ferromagnetic state which is a zero-energy state of the $c_2$ term, as it does not contain any spin singlets. 
Therefore, Eq.~(\ref{Thm1}) is 
an eigenstate of $\hat{H}_b$. 
Since $\hat{S}_{i, +}\hat{S}_{i, -}$ is positive semidefinite, we have $\hat{H}_a \leqslant \hat{H}_b$.
Then the min-max theorem implies that $E_{a,0}= E_{b,0}$ and $E_{a,1} \leqslant E_{b,1}$. Recalling that $E_{a,0}< E_{a,1}$, we get $E_{b,0}<E_{b,1}$. This proves that the local ground state in the case $c_1<0$ and $c_2>0$ is also unique.

\smallskip

\textit{Proof of Proposition 1.---}Under the conditions of Proposition~1, the Hamiltonian in Eq.~(\ref{Hamiltonian}) becomes
$\hat{H}'=-\sum_{ i\neq j, \alpha} t_{i, j} \hat{a}^{\dagger}_{i,\alpha} \hat{a}_{j,\alpha} + \sum_{i}V_i \hat{n}_i 
+(c_0/8)\sum_{i}[2\hat{n}_i(2\hat{n}_i+1)- (\hat{\bm{F}_i})^2]
+ (2c_2/5) \sum_{i}\hat{S}_{i, +}\hat{S}_{i, -}$.
We seek states (if any) that minimize all terms in $\hat{H}'$ simultaneously.
The last two terms in $\hat{H}'$ are now positive semidefinite.
According to Theorem~1, the ground states must have $F_{\text{tot}}=2N$, which clearly makes the last two terms zero.
As for the hopping term, according to the Perron-Frobenius theorem,
its single-particle ground state $\varphi_0$ is unique and
satisfies $\varphi_0(i)>0$.
Thus it is obvious that $(\hat{b}^{\dagger}_{2})^N |\text{vac}\rangle$
in Eq.~(\ref{Prop1}) gives the unique ground state in $\mathcal{H}_{M=2N}$.
Since $[\hat{H}, \hat{F}_{\text{tot}}^-]$, the uniqueness of local ground states
then implies $|\Psi^{\text{GS}}_{M'}\rangle \propto (\hat{F}_{\text{tot}}^-)^{2N-M'} |\Psi^{\text{GS}}_{M=2N}\rangle$.

\smallskip


\textit{Proof of Theorem 2.---}There exists a new set of bosonic operators $\hat{d}^{\dagger}$ and $\hat{d}$ defined in the Supplemental Material~[\hyperref[footnote1]{36}],
such that the singlet creation operator can be written as
$\hat{S}_{i,+} = (\sum^5_{\mu=1}\hat{d}^{\dagger}_{i,\mu} \hat{d}^{\dagger}_{i,\mu})/2$.
The form of $\hat{S}_{i,+}$ now remains unchanged
when $\hat{d}^{\dagger}_{i,\mu}$'s are subject to SO(5) transformations. 
Thus the $c_2$ term has a manifest SO(5) symmetry. 
In the Cartan-Weyl basis, ten generators of SO(5) are
\begin{equation}
\hat{E}_{i,\alpha\beta} = (-1)^\alpha \hat{a}^{\dagger}_{i,\alpha} \hat{a}_{i,-\beta} - (-1)^\beta \hat{a}^{\dagger}_{i,\beta} \hat{a}_{i,-\alpha}, \label{generator}
\end{equation}
where $-2\leqslant \beta<\alpha\leqslant2$.
Taking $\alpha=-\beta$, we get a basis of the Cartan subalgebra:
$\hat{P}_i := \hat{E}_{i,-1,1} = \hat{n}_{i,1} - \hat{n}_{i,-1}$
and $\hat{Q}_i := \hat{E}_{i,2,-2} = \hat{n}_{i,2} - \hat{n}_{i,-2}$.
Indices $(\alpha,\beta)$ of the other eight generators
are roots in the root system $B_2$. 

\smallskip

Because of the SO(5) symmetry of the Hamiltonian, 
we have $[\hat{P},\hat{H}] = [\hat{Q},\hat{H}] = 0$,
which shows that $\hat{H}$ conserves $P=N_1-N_{-1}$ and $Q=N_2-N_{-2}$
and splits into blocks with respect to these two quantum numbers.
Besides the hopping term, off-diagonal matrix elements appear only in the $c_2$ term.
By applying the following U(1) transformation for every $i\in \Lambda$:
$\hat{\tilde{a}}^{\dagger}_{i,\pm1} = \mathrm{i}\hat{a}^{\dagger}_{i,\pm1}$,
$\hat{\tilde{a}}^{\dagger}_{i,\pm2} = \hat{a}^{\dagger}_{i,\pm2}$ and
$\hat{\tilde{a}}^{\dagger}_{i,0} = \hat{a}^{\dagger}_{i,0}$,
one can verify that all the off-diagonal matrix elements of $\hat{H}$ in this basis become nonpositive. 
Furthermore, connectivity of configurations and that of spin states are guaranteed by the hopping term and the $c_2$ term, respectively. 
The Perron-Frobenius theorem is thus applicable and asserts that the local ground state $| \Psi^{\text{GS}}_{P,Q} \rangle$ in $\mathcal{H}_{P,Q}$ is unique and can be written as Eq.~(\ref{Thm2}). 

\smallskip

Now we extract some useful information from the aforementioned single-site model. 
We claim that states written in the form 
$|(\bm{\alpha}, \bm{\beta});N,v\rangle := \hat{E}_{q,\alpha_m\beta_m} \cdots \hat{E}_{q,\alpha_2\beta_2} \hat{E}_{q,\alpha_1\beta_1} (\hat{S}_{q,+})^{(N-v)/2}(a^{\dagger}_{q,+2})^{v} |\text{vac}\rangle$ 
are eigenstates of $\hat{S}_{q,+} \hat{S}_{q,-}$ with eigenvalues $(N^2+3N-v^2-3v)/4$~[\hyperref[footnote1]{36}].
Since $c_2<0$, the smaller $v$, the lower energy.
Define $\Gamma:=|P|+|Q|$ and note that $v \geqslant \Gamma\geqslant 0$.
Since $v$ is the number of particles that do not form singlets, 
for a ground state of the single-site model $| \widetilde{\Psi}^{\text{GS}}_{P,Q} \rangle$ 
$v$ takes the minimum possible value, which is $v_{\text{min}} = \Gamma$ if $N-\Gamma$ is even, while $v_{\text{min}} = \Gamma +1$ if $N-\Gamma$ is odd.
The Casimir operator for the model on the total lattice $\Lambda$ is defined as $\hat{C}_\text{tot}^2 = \sum_{\alpha < \beta} ( \hat{E}_{\alpha\beta}^\dagger \hat{E}_{\alpha\beta} + \hat{E}_{\alpha\beta} \hat{E}_{\alpha\beta}^\dagger)/2$, where $\hat{E}_{\alpha\beta} := \sum_i \hat{E}_{i,\alpha\beta}$. It is easy to see that $\hat{C}_\text{tot}^2 | \widetilde{\Psi}^{\text{GS}}_{P,Q} \rangle = [\hat{N}(\hat{N}+3)-4\hat{S}_{q,+} \hat{S}_{q,-}] | \widetilde{\Psi}^{\text{GS}}_{P,Q} \rangle = v_{\text{min}}(v_{\text{min}}+3) | \widetilde{\Psi}^{\text{GS}}_{P,Q} \rangle$. 
Since the Perron-Frobenius theorem again applies to the single-site model,
we have $\langle \widetilde{\Psi}^{\text{GS}}_{P,Q} | \Psi^{\text{GS}}_{P,Q} \rangle \neq 0$, 
which implies
\begin{eqnarray}
	\hat{C}_\text{tot}^2  | \Psi^{\text{GS}}_{P,Q} \rangle 
	= v_{\text{min}}(v_{\text{min}}+3) | \Psi^{\text{GS}}_{P,Q} \rangle.
\end{eqnarray}

We are now ready to prove the energy-level ordering. Let $E^{\text{GS}}_{P,Q}$ be the energy of the local ground state $|\Psi^{\text{GS}}_{P,Q} \rangle$. We first note that $E^{\text{GS}}_{P,Q}=E^{\text{GS}}_{|P|,|Q|}$,
because 
under the transformation $\hat{a}_{+2} \leftrightarrow \hat{a}_{-2}$ or $\hat{a}_{+1} \leftrightarrow \hat{a}_{-1}$, the Hamiltonian remains unchanged 
but $\hat{P}$ or $\hat{Q}$ gets a minus sign.
Thus it suffices to consider the case $P,Q \geqslant 0$. 
Next, we prove that all $E^{\text{GS}}_{P,Q}$'s with the same $\Gamma$ are the same. 
Define $|\Psi_a \rangle= \hat{E}_{2,-1} |\Psi_{P+1,Q-1}^{\text{GS}}\rangle \in \mathcal{H}_{P,Q}$ (assume $Q\geqslant 1$).
Apparently energy eigenvalue of $|\Psi_a\rangle$ should be the same as $|\Psi_{P+1,Q-1}^{\text{GS}}\rangle$, which is $E^{\text{GS}}_{P+1,Q-1}$. 
So we have $E^{\text{GS}}_{P,Q} \leqslant E^{\text{GS}}_{P+1,Q-1}$.
Define $|\Psi_b \rangle= \hat{E}_{1,-2} |\Psi_{P,Q}^{\text{GS}}\rangle \in \mathcal{H}_{P+1,Q-1}$ (assume $Q\geqslant 1$),
and similarly we get $E^{\text{GS}}_{P+1,Q-1} \leqslant E^{\text{GS}}_{P,Q}$.
Thus, we have $E^{\text{GS}}_{P+1,Q-1} = E^{\text{GS}}_{P,Q}$, which means that  $E^{\text{GS}}_{P,Q}$ is only a function of $\Gamma=|P|+|Q|$,
denoted as $E^{\text{GS}}_{\Gamma}$. 
Now we show the ordering of $E^{\text{GS}}_{\Gamma}$. 
Construct $|\Psi_c \rangle = \hat{E}_{0,-1} |\Psi_{P+1,Q}^{\text{GS}}\rangle \in \mathcal{H}_{P,Q}$ and then get $E^{\text{GS}}_{\Gamma} \leqslant E_{\Gamma+1}^{\text{GS}}$.
When $N-\Gamma$ is even, $|\Psi_c\rangle$ and $|\Psi_{P,Q}^{\text{GS}}\rangle$ have different $C_{\text{tot}}^2$, and hence are orthogonal. 
The uniqueness of each local ground state then yields $E^{\text{GS}}_{\Gamma} < E^{\text{GS}}_{\Gamma+1}$.
When $N-\Gamma$ is odd, construct $|\Psi_d\rangle = \hat{E}_{1,0} |\Psi_{P,Q}^{\text{GS}}\rangle \in \mathcal{H}_{P+1,Q}$, 
and similarly we have $E^{\text{GS}}_{\Gamma} \geqslant E^{\text{GS}}_{\Gamma+1}$, 
which finally gives $E^{\text{GS}}_{\Gamma} = E^{\text{GS}}_{\Gamma+1}$. 
We thus have obtained the desired energy-level ordering stated in \textit{Theorem 2}. 
Consequently, the global ground state is unique and lies in the subspace $\mathcal{H}_{\Gamma=0}$ when $N$ is even, while it is five-fold degenerate and lies in $\mathcal{H}_{\Gamma=0} \oplus \mathcal{H}_{\Gamma=1}$ when $N$ is odd. 
Then it follows from $[\hat{H}, (\hat{\bm{F}}_{\text{tot}})^2]=0$ that the global ground state has $F_{\text{tot}}=0$ ($F_{\text{tot}}=2$) when $N$ is even (odd). 

\smallskip

\textit{Proof of Theorem 3.---}Applying the Perron-Frobenius theorem to the hopping term proves the theorem. 
The proof is essentially the same as that of Theorem 3 in \cite{Katsura2013} (see the Supplemental Material~[\hyperref[footnote1]{36}] for details).



\smallskip

{\it Discussion.}---In conclusion, we have established the basic ground-state properties of the spin-2 Bose-Hubbard model, as stated in the main theorems. Symmetry plays an important role in our proofs. In particular, the SO(5) symmetry is essential in the case $\{ c_1 = 0, c_2 < 0 \}$.  
Although the Cartan subalgebra of $\mathfrak{so}(5)$ is two-dimensional, we found that the energy-level ordering is effectively ``one-dimensional", as it is characterized only by the quantum number $v$. 

\smallskip

In the presence of an external magnetic field in the $z$ direction,
one should add linear and quadratic Zeeman terms
$\sum_{i,\alpha} (-p_i \alpha \hat{n}_{i,\alpha}+ q_i \alpha^2 \hat{n}_{i,\alpha})$ to $\hat{H}$~\cite{Ueda2002,Uchino2010}.
In this case, the total Hamiltonian no longer has SO(3) symmetry. 
However, since the Zeeman terms are diagonal in the basis Eq.~(\ref{basis}),
the uniqueness of the ground state within each subspace as well as Eqs.~(\ref{Thm1})--(\ref{Thm3}) still holds in each respective parameter region. 

\smallskip

\begin{acknowledgments}
The authors would like to thank Masahito Ueda and Sho Higashikawa for fruitful discussions. H. K. was supported in part by JSPS KAKENHI Grants No. JP15K17719, No. JP16H00985, No. JP18K03445, and No. JP18H04478.
\end{acknowledgments}

\bibliography{Manuscript_Spin-2}

\widetext

\vspace{20pt}

\begin{center}
	\textbf{\large Supplemental Material of \\
``\textit{Rigorous Results for the Ground States of the Spin-2 Bose-Hubbard Model}" }
\end{center}

\newtagform{supplemental}{(S}{)}
\usetagform{supplemental}
\setcounter{equation}{0}

\section{SO(5) symmetry of singlet-creation operator}
Here we show how to express the singlet-creation 
operator in the form with explicit SO(5) symmetry.
In the so-called $d$-orbital basis that is obtained by the following unitary transformation 
\begin{equation}
\renewcommand\arraystretch{1.2}
\left(
\begin{array}{l}
\hat{d}^{\dagger}_{i,1} \\ \hat{d}^{\dagger}_{i,2} \\ \hat{d}^{\dagger}_{i,3} \\ \hat{d}^{\dagger}_{i,4} \\ \hat{d}^{\dagger}_{i,5}
\end{array}\right)
:=\frac{1}{\sqrt{2}}
\begin{pmatrix}
\mathrm{i} &0 &0 &0 &{-\mathrm{i}} \\
0 &\mathrm{i} &0 &\mathrm{i} &0 \\
0 &0 &\sqrt{2} &0 &0 \\
0 &1 &0 &{-1} &0 \\
1 &0 &0 &0 &1
\end{pmatrix}
\left(
\begin{array}{l}
\hat{a}^{\dagger}_{i,-2} \\ \hat{a}^{\dagger}_{i,-1} \\ \hat{a}^{\dagger}_{i,0} \\ \hat{a}^{\dagger}_{i,1}\\ \hat{a}^{\dagger}_{i,2}
\end{array}\right),
\end{equation}
singlet creation operator can be written as
\begin{equation}
	\hat{S}_{i,+} =\frac{1}{2}\sum_{\alpha=-2}^2 (-1)^{\alpha}{\hat a}^{\dagger}_{i,\alpha} {\hat a}^{\dagger}_{i,-\alpha}= \frac{1}{2}\sum^5_{\mu=1}\hat{d}^{\dagger}_{i,\mu} \hat{d}^{\dagger}_{i,\mu}.
\end{equation}
The SO(5) symmetry now becomes clear in $d$-orbital basis.

\section{Eigenstates of $\hat{S}_{q,+} \hat{S}_{q,-}$ from highest-weight representation of $\frak{so}(5)$}

In this section we discuss how to find
all the eigenstates and eigenvalues of $\hat{S}_{q,+} \hat{S}_{q,-}$
using the highest-weight representation of $\frak{so}(5)$.
The Cartan-Weyl basis of $\frak{so}(5)$ algebra at site $q$ is
\begin{equation}
\hat{E}_{q,\alpha\beta} = (-1)^\alpha \hat{a}^{\dagger}_{q,\alpha} \hat{a}_{q,-\beta} - (-1)^\beta \hat{a}^{\dagger}_{q,\beta} \hat{a}_{q,-\alpha} \label{generator}
\end{equation}
with $(\alpha,\beta) \in B_2$.
Two simple roots in $B_2$ are $\Delta=\{(2,-1),(1,0)\}$.
A highest weight state of $\frak{so}(5)$ algebra at site $q$ is 
an eigenstate of Cartan subalgebra ($\hat{P}_q$ and $\hat{Q}_q$)
that is annihilated by $\hat{E}_{q,\alpha\beta}$ with $(\alpha, \beta) \in \Delta$.
Thus, the only way to construct a highest-weight state is
\begin{equation}
	|N,v\rangle := (\hat{S}_{q,+})^{(N-v)/2}(a^{\dagger}_{q,+2})^{v} |\text{vac}\rangle.
\end{equation}
Define states generated from the highest-weight state as 
\begin{equation}
	|(\bm{\alpha}, \bm{\beta});N,v\rangle := \hat{E}_{q,\alpha_m\beta_m} \cdots \hat{E}_{q,\alpha_2\beta_2}\hat{E}_{q,\alpha_1\beta_1} |N,v\rangle, \label{descender}
\end{equation}
where $(\alpha_1,\beta_1),...,(\alpha_m,\beta_m) \in -\Delta = \{(1,-2),(0,-1)\}$ and $m$ can be any non-negative integer 
as long as $|(\bm{\alpha}, \bm{\beta});N,v\rangle \neq 0$. 
We claim that $|(\bm{\alpha}, \bm{\beta});N,v\rangle$ is an eigenstate of $\hat{S}_{q,+} \hat{S}_{q,-}$, with eigenequation
\begin{equation}
	\hat{S}_{q,+} \hat{S}_{q,-} |(\bm{\alpha}, \bm{\beta});N,v\rangle 
 = \frac{1}{4}(N^2+3N-v^2-3v) |(\bm{\alpha}, \bm{\beta});N,v\rangle.
\end{equation}
To prove it, recall that the SO(5) symmetry leads to $[ \hat{E}_{q,\alpha\beta}, \hat{S}_{q,+} ]=0$,
and the desired result then follows from an iterated application of the identity: $[ \hat{S}_{q,-}, \hat{S}_{q,+} ] = (2\hat{N}+5)/2$. 
Note that for a given $(\bm{\alpha},\bm{\beta})$, the eigenstate $|(\bm{\alpha}, \bm{\beta});N,v\rangle$ 
can be labeled by quantum numbers $(N,v,P_q,Q_q)$. The state is, however, not necessarily an eigenstate of $({\hat {\bm F}}_q)^2$, 
which is in contrast to the eigenstates constructed in \cite{Ueda2002}, labeled by $(N,v,F_{q},F_q^z)$. 

Now we are going to show that all the eigenstates have been found by the construction of Eq.~(\hyperref[descender]{S5}).
With fixed $(N,v)$, $[N,v]:=\text{span}(\{ |(\bm{\alpha}, \bm{\beta});N,v\rangle \})$
forms a highest-weight representation space (module) of $\frak{so}(5)$~\cite{georgi1999lie}.
For $N$ spin-2 bosons on the same site $q$, the Hilbert space must be symmetric.
This symmetric space can be decomposed as
\begin{equation}
	(\overbrace{\mathcal{H}^5\otimes\mathcal{H}^5\otimes\cdots\otimes\mathcal{H}^5}^{N} )_{\text{sym}}
 =[N,v=N]\oplus[N,v=N-2]\oplus\cdots\oplus[N,v=1] \text{ or }[N,v=0],
\end{equation}
where $\mathcal{H}^5$ is the five-dimensional Hilbert space of a single spin-2 particle. 
Each subspace denoted by $[N,v]$ corresponds to an eigenspace of $\hat{S}_{q,+} \hat{S}_{q,-}$. 
We thus have already found all eigenstates of $\hat{S}_{q,+} \hat{S}_{q,-}$,
i.e., all eigenstates can be expressed as linear combinations of the states $|(\bm{\alpha}, \bm{\beta});N,v\rangle$. 
This decomposition is nothing but a way of constructing irreducible representation of SO(5) group~\cite{zee2016group}.

\section{On the proof of Theorem 3 }

Proof of Theorem 3 is rather straightforward. However, readers who may need further details may refer to the following.
In the case of Theorem 3, it suffices to consider the subspaces labeled by $\{ N_\alpha \}^5_{\alpha=1}$ separately. 
It is easy to see that all possible states in each subspace are connected via the off-diagonal elements $\langle \Phi_{\bm{m}} | \hat{H} | \Phi_{\bm{m}'} \rangle \leqslant 0$ (${\bm m} \ne {\bm m}'$), which result from the hopping term. 
Then the Perron-Frobenius theorem guarantees that the ground state within each $\mathcal{H}_{ N_{-2},...,N_2}$ is unique and is written as Eq.~(6). 
The ground-state degeneracy is exactly the same as the number of subspaces.

\end{document}